\documentclass[12pt]{article}
\usepackage{geometry}
\usepackage{amsmath}
\usepackage{amssymb,epsfig,subfigure}
\usepackage{graphicx}
\usepackage{cite}
\numberwithin{equation}{section}

\newcommand{\be}{\begin{equation}}
\newcommand{\ee}{\end{equation}}
\newcommand{\ba}{\begin{aligned}}
\newcommand{\ea}{\end{aligned}}

\def\m1{\left(-1\right)^{F_i}}

\makeatletter
\def\sla@#1#2#3#4#5{{%
  \setbox\z@\hbox{$\m@th#4#5$}%
  \setbox\tw@\hbox{$\m@th#4#1$}%
  \dimen4\wd\ifdim\wd\z@<\wd\tw@\tw@\else\z@\fi
  \dimen@\ht\tw@
  \advance\dimen@-\dp\tw@
  \advance\dimen@-\ht\z@
  \advance\dimen@\dp\z@
  \divide\dimen@\tw@
  \advance\dimen@-#3\ht\tw@
  \advance\dimen@-#3\dp\tw@
  \dimen@ii#2\wd\z@  \raise-\dimen@\hbox to\dimen4{%
    \hss\kern\dimen@ii\box\tw@\kern-\dimen@ii\hss}%
  \llap{\hbox to\dimen4{\hss\box\z@\hss}}}}
\def\slashed#1{%
  \expandafter\ifx\csname sla@\string#1\endcsname\relax
    {\mathpalette{\sla@/00}{#1}}%
  \else
    \csname sla@\string#1\endcsname
  \fi}
\makeatother

\newcommand\e{\mathrm{e}}

\newcommand\nn{\nonumber}

\newcommand{\sch}{Schr\"odinger }

\begin{document}


\thispagestyle{empty}
\begin{flushright}\footnotesize
\texttt{CALT-68-2724}\\
\texttt{IPMU09-0036, UT-09-09} \\
\texttt{KUNS-2198}
\vspace{2cm}
\end{flushright}

\renewcommand{\thefootnote}{\fnsymbol{footnote}}
\setcounter{footnote}{0}

\begin{center}
{\Large\textbf{\mathversion{bold} Coset Construction for Duals of Non-relativistic CFTs}\par}

\vspace{2cm}

\textrm{Sakura Sch\"afer-Nameki$^1$, Masahito Yamazaki$^{1,2}$,  and Kentaroh Yoshida$^3$}

\vspace{1cm}

$^1$\textit{California Institute of Technology\\
1200 E California Blvd., Pasadena, CA 91125, USA } \\
\vspace{0.5cm}

$^2$\textit{Institute for the Physics and Mathematics of the Universe, \\
University of Tokyo, Kashiwa, Chiba 277-8596, Japan \\
and\\
Department of Physics, University of Tokyo,\\ 
Hongo 7-3-1, Tokyo, 113-0033, Japan} 
\vspace{0.5cm}

$^{3}${\it Department of Physics, Kyoto University \\ 
Kyoto 606-8502, Japan} 
\vspace{0.5cm}

{\tt ss299 at theory.caltech.edu, } \ 
{\tt yamazaki at hep-th.phys.s.u-tokyo.ac.jp}  \\
{\tt kyoshida at gauge.scphys.kyoto-u.ac.jp}

\bigskip


\par\vspace{1cm}

\textbf{Abstract}\vspace{5mm}
\end{center}

\noindent
We systematically analyze backgrounds that are holographic duals to non-relativistic CFTs, by constructing them as cosets of the Schr\"odinger group and variants thereof. These cosets $G/H$ are generically non-reductive and we discuss in generality how a metric on such spaces can be determined from a non-degenerate $H$-invariant symmetric two-form. Applying this to the $d=2$ Schr\"odinger algebra, we reproduce the five-dimensional backgrounds proposed as duals of fermions at unitarity, and under reasonable physical assumptions, we demonstrate uniqueness of this background. 
The proposed gravity dual of the Lifshitz fixed-point, for which Galileian symmetry is absent, also fits into this organizational scheme and uniqueness of this background can also be shown.

\vspace*{\fill}

\setcounter{page}{1}
\renewcommand{\thefootnote}{\arabic{footnote}}
\setcounter{footnote}{0}

 \newpage




\section{Introduction and Summary}

Variants of the AdS/CFT correspondence, which provide gravity duals of non-relativistic gauge theories, could potentially be of great importance in order to describe strongly coupled, scale-invariant condensed matter systems. Examples of such systems are fermions at unitarity and theories at Lifshitz-like fixed points.

The simplest proposals for gravity duals to non-relativistic theories were put forward in 
\cite{Son:2008ye, Balasubramanian:2008dm, Kachru:2008yh}. One natural question that arises is, whether these spaces are homogeneous and whether they are a comprehensive list of such backgrounds.
In this note, we want to address this question, and find an organizational principle to study duals to non-relativistic theories as homogeneous  spaces. The particular cosets involved in this construction are rather non-standard, in that they do not  generically yield symmetric spaces. We provide a framework for studying such backgrounds and then demonstrate that under certain physical assumptions, the metrics found in \cite{Son:2008ye, Balasubramanian:2008dm, Kachru:2008yh} are unique. It will be interesting to extend this to the supercosets for the super-Schr\"odinger algebras and study the corresponding backgrounds for superstring theory. 

We begin in section \ref{sec:General} by discussing the general theory of invariant metrics on cosets, in particular focusing on their existence for general cosets that are not necessarily reductive \footnote{For discussions of reductive cosets for another non-relativistic algebra (Newton-Hooke algebra), 
see \cite{BGK,AGKP}.}. This will be important, as the background found in \cite{Son:2008ye,  Balasubramanian:2008dm} are non-reductive cosets of the Schr\"odinger algebra \cite{Sch1,Sch2}.
In order to construct the metric on these cosets, the key ingredient is the existence of a nondegenerate symmetric two-form, that is invariant under the denominator group, as in \cite{NW}. We apply this general theory to spaces with Schr\"odinger symmetry in section  \ref{sec:Sch} and to the dual of the Lifshitz fixed point in section \ref{sec:Lif}, and demonstrate how these are unique under certain physical assumptions on the subgroup.


\section{General Considerations on Cosets}
\label{sec:General}


\subsection{Homogeneous Spaces and Invariant Two-forms}

Consider a coset (homogeneous space) $M=G/H$, where $G$ is a Lie group and $H$ is a Lie subgroup of $G$. Let us denote the corresponding Lie algebras by $\mathfrak{g}$ and $\mathfrak{h}$, respectively. For each $g\in \mathfrak{g}$, let us denote the corresponding element of $\mathfrak{g}/\mathfrak{h}$ by $[g]$.
As a vector space, we can always decompose 
\be
\mathfrak{g} = \mathfrak{h} \oplus \mathfrak{m} \,,  \label{eq.g=h+m}
\ee
but there is an ambiguity in the choice of $\mathfrak{m}$. 
One can impose various compatibility conditions of the Lie algebra structure with this linear space decomposition. The coset $M$ is called a reductive coset if there is a choice of $\mathfrak{m}$ such that it is ad$(\mathfrak{h})$-invariant, i.e.
\be
[\mathfrak{h}, \mathfrak{m}] \subset  \mathfrak{m} \label{HMM}  \,.
\ee
If we impose in addition that 
\be
[\mathfrak{m}, \mathfrak{m}] \subset  \mathfrak{h} \label{MMH} \,,
\ee
then $M$ is a symmetric space, which is equivalent to the existence of a $\mathbb{Z}_2$ grading, such that deg$(\mathfrak{h})=0$, deg$(\mathfrak{m})=1$, which is compatible with the Lie algebra structure. 

We wish to construct a $G$-invariant metric on the homogenous space $M$. 
If $\mathfrak{g}$ is semi-simple, then the Killing form is non-degenerate and induces a $G$-invariant metric on $M$.
In the case of degenerate Killing form, the existence of such a $G$-invariant metric is not guaranteed, however the following proposition gives a useful criterion:

\medskip
\noindent
\textbf{Proposition} (\cite{KN2}, Proposition X.3.1)\\
There is a one-to-one correspondence between $G$-invariant indefinite Riemannian metrics $\mathcal{G}$ on $M= G/H$ and Ad$(H)$-invariant non-degenerate symmetric bilinear forms $\Omega$ on $\mathfrak{g}/\mathfrak{h}$. When $H$ is connected, Ad$(H)$-invariance of $\Omega$ reduces to ad$(\mathfrak{h})$-invariance, meaning that
\be
\Omega ([h , [t_1]], [t_2]) + \Omega([t_1], [h, [t_2]]) = 0\,,\label{eq.Hinv}
\ee
for any $h\in \mathfrak{h}$, or equivalently, written in terms the structure constants of the Lie algebra,
\be\label{TwoForm}
\Omega_{[m][n]} f_{[k]p}\,^{[m]} + \Omega_{[k][m]} f_{[n]p}\,^{[m]} =0 \,, 
\ee
where $[m], [n]\cdots $ denotes generators of $\mathfrak{m}$ and $p$ is an $\mathfrak{h}$-index. The structure constants  $f_{[k]p}\,^{[m]}$ are well-defined since $\mathfrak{h}$ is a subalgebra and different representatives of the coset element $[k]$ give the same answer.

\medskip

For our purposes, it is important to know the explicit relation between $\Omega$ and the metric. Let us first define a metric on the identity element $[e]=eH$ of $M=G/H$. We want to define a metric $\mathcal{G}(X_1,X_2)$, where $X_1$ and $X_2$ are elements of $T_e M$. Recall that $T_eG=\mathfrak{g}$ and likewise, $T_eM=\mathfrak{g}/\mathfrak{h}$. Therefore, $X_1$ and $X_2$ can be identified with elements $[t_1]$ and $[t_2]$ of $\mathfrak{g}/\mathfrak{h}$, respectively. Under this identification, the explicit correspondence between $\Omega$ and $\mathcal{G}$ is given by
\be
\mathcal{G}(X_1,X_2)_{[e]}=\mathcal{G}([t_1],[t_2])_{[e]}=\Omega([t_1],[t_2]) \,,\qquad  t_1\,, \ t_2\in \mathfrak{g} \,. \label{eq.Omega=g}
\ee

Next we need to define a metric at an arbitrary point $[g]$ of $M=G/H$. By the $G$-action on the whole manifold, it is possible to translate the metric at the origin to any other point. Choose an arbitrary representative $g$ of $[g]$. Then left-multiplication by $g^{-1}$ yields the map $L_{g^{-1}}:M \rightarrow M$.
From this map we have an induced map 
\be
[J_g]:=(L_{g^{-1}})_*:\ T_{[g]}M\rightarrow T_{[e]} M=\mathfrak{g}/\mathfrak{h}\,,
\ee
which is a $\mathfrak{g}/\mathfrak{h}$-valued one-form on $\mathfrak{m}$. This is the Maurer-Cartan (MC) one-form. Similarly, we can define $J_g:T_{g}G\rightarrow T_eG=\mathfrak{g}$. 

For two vector fields $Y_1,Y_2\in T_{[g]}M$, define the metric $\mathcal{G}(Y_1,Y_2)_{[g]}$ at point $[g]$ to be
\be
\mathcal{G}(Y_1,Y_2)_{[g]}=\mathcal{G}([J_g] (Y_1),[J_g] (Y_2))_{[e]}\,.\label{eq.G_g}
\ee
When $\mathfrak{g}$ is embedded into $\mathfrak{gl}_N$ and $g$ takes matrix values (which we will assume throughout this paper), the MC one-form can be written as 
\be
[J_g]=[g^{-1}dg] \,. \label{eq.MC}
\ee
In the discussion above we chose a particular representative $g$ for $[g]$. If we choose another representative $gh$ with $h\in H$, we have  
\be
[J_{gh}]=[(gh)^{-1} d(gh)]=[{\rm ad}(h)(g^{-1} dg)],
\ee
where we have used the relation $[h^{-1} dh]=0$ as an element of $\mathfrak{g}/\mathfrak{h}$. This relation, together with the definition of the metric in \eqref{eq.G_g} and the $H$-invariance condition of \eqref{eq.Hinv}, tells us that the metric \eqref{eq.G_g} is independent of the choice of the representative of $[g]$. This shows the well-definedness of the metric.

\medskip

Several comments are now in order:

\begin{enumerate}

\item
The structure constants $f_{[k]p}\,^{[m]}$ we used above are in general different from the structure constants $f_{kp}\,^{m}$ of $\mathfrak{g}$. They are equivalent only for reductive cosets \eqref{HMM}. 

\item Whenever $\mathfrak{g}$ is semi-simple, the Killing form is non-degenerate and provides a natural candidate for $\Omega$.
In many instances that will be of interest to us, the Killing form is degenerate and the invariant non-degenerate two-form we use is different from the Killing form.

\item

Reductiveness is a natural notion for Riemannian cosets: if $G$ is an isometry group of a Riemannian metric on $G/H$ and if $H$ is connected, 
then $G/H$ is automatically reductive \cite{KN2}. However, in Lorentzian signature this
is in general not true and some of the examples we discuss below are indeed non-reductive.
We therefore do not impose either condition \eqref{HMM} and \eqref{MMH} in the following discussions.

\item We emphasize again that in general neither existence nor uniqueness of such a two-form $\Omega$ is guaranteed. For some coset $G/H$, $\Omega$ does not exist, and for others there exists a family of such invariant two-forms, as we shall see exemplified below.

\item A homogeneous space is mathematically defined as a space $M$ with a transitive action of $G$, meaning for any two points $x,x'$ of $M$ we can find an element $g_{x,x'}$ of $G$ such that $g_{x,x'}.x=x'$. From this condition it follows that
$M$ is written as a coset $G/G_x$, where $G_x$ is the stabilizer at point $x$.
If we choose a different point $x'$, $G_{x'}$ and $G_x$ are in general different, but belong to the same conjugacy class. 
This means that classification of cosets of the form $G/H$ for a given $G$ reduces to the two problems: first to classify conjugacy classes of its subgroups and second to classify the non-degenerate invariant two-forms of the subgroup. 


\end{enumerate}

In summary, a homogeneous space is characterized by the data $(G, H, \Omega)$, where $\Omega$ is a $\mathfrak{h}$-invariant nondegenerate symmetric two-form specifying the $G$-invariant metric on the coset space $G/H$. 
We will apply this general discussion to the cases of interest in the context of non-relativistic conformal theories.


\subsection{Explicit Coordinate Description of Cosets}

In the previous section, we used a coordinate invariant formalism. However, in order to derive explicit forms of the metrics it is often useful to go to a particular coordinate frame. 
For that purpose, we first fix a linear space decomposition \eqref{eq.g=h+m}, as well as  a basis $t_m, t_n,...$ for $\mathfrak{h}$ and $t_p, t_q,...$ for $\mathfrak{m}$. Then we parametrize an element $[g]\in G/H$ by 
\be
[g]=[\exp(x_m t_m) \exp(x_n t_n).....] \quad (\textrm{modulo } H). \label{eq.grep}
\ee
Of course, the expression in \eqref{eq.grep} is far from unique. For example, another possible parametrization is
\be
[g]=[\exp(\sum_{m,n,..} x_m t_m)].
\ee
These are just different choices of coordinates on $G/H$ and are related by coordinate transformations. We will thus choose a convenient expression in each of the subsequent discussions.

The MC one-form $J_g=g^{-1} dg$ can now be computed explicitly and decomposed according to \eqref{eq.g=h+m}:
\be
J_g=e_m t_m+e_p t_p.\label{eq.Jdecomp}
\ee
In this notation, the metric defined in \eqref{eq.Omega=g},\eqref{eq.G_g} is written as
\be
G=\Omega^{mn} e_m e_n\,,
\ee
namely, $e_m$ are nothing but vielbeine, which get contracted with $\Omega$. 
%
If we choose a different representation $gh$ for $[g]$, 
\be
J_{gh}=\textrm{Ad}(h) (J_g),
\ee
and the vielbeine mix among themselves, which shows that $H$ is a symmetry of the vielbeine. 


\section{The Schr\"odinger Algebra and Cosets}
\label{sec:Sch}

\subsection{The Schr\"odinger Algebra}

The Schr\"odinger algebra $\mathfrak{Sch}_d$ in $d+1$ dimensions    \cite{Sch1,Sch2} has generators $J^{ij}$ (spatial rotations), $P^i$ (spatial translations), $H$ (Hamiltonian), $G^i$ (Galileian boosts), $D$ (dilatation) and $C$ (special conformal transformations). One can consider a central extension $\widetilde{\mathfrak{Sch}_d}$ of this algebra by the mass operator $M$. The 
non-vanishing commutation relations are
\begin{eqnarray}
&& [J^{ij},J^{kl}]=-\delta^{ik} J^{jl}+\delta^{il} J^{jk} - \delta^{jl}J^{ik} + \delta^{jk} J^{il}\,, 
\nonumber \\ 
&& [J^{ij},P^k] = -\delta^{ik}P^j + \delta^{jk}P^i\,, \qquad 
[J^{ij},G^k] 	= -\delta^{ik} G^j + \delta^{jk}G^i\,, \nonumber \\ 
&& [H,G^i] 	=P^i    \,,\qquad\quad   [D,G^i] 		= -G^i\,, \qquad 
[C,P^i] 	=-G^i   \,,\qquad    [D,P^i]		 =P^i\,, \nonumber \\ 
 && [D,H]	=2H \,, \phantom{+}\qquad [H,C] =-D\,, \qquad  [D,C]=-2C\,,
\end{eqnarray}
as well as the central extension
\be
 [P^i,G^j]  = \delta^{ij}M    \,.
\ee
This algebra is a subalgebra of conformal algebra, as was observed in \cite{Son:2008ye, Sakaguchi:2008rx, Sakaguchi:2008ku}.


\subsection{Subalgebras and Two-forms for $d=2$}\label{sec.d=3}

Let us consider the case $d=2$. In this case, denoting $J^{12}=J$, the algebra is \footnote{We use the same symbole $H$ to denote the Hamiltonian of $\mathfrak{Sch}$ and the denominator subgroup of the coset. We hope no confusion will arise.}
\begin{eqnarray}
&& [J,P^1] 	= - P^2 \,,\qquad   [J,P^2] 	=  P^1\,, \qquad 
[J,G^1]  = -G^2\,,\qquad     [J,G^2] 	= G^1\,,  \nonumber \\
&& [H,G^i] =   P^i\,, \phantom{+}\qquad  [D,G^i] 		= -G^i\,, \qquad 
[C,P^i] = -G^i\,, \qquad    [D,P^i]		 =P^i\,, \nonumber \\
&&  [D,H]   =2H\,, \phantom{+}\qquad [H,C] =-D\,, \qquad  [D,C]=-2C\,, \qquad 
[P^i,G^j] = \delta^{ij}M    \,.
\end{eqnarray}
For $d=3$ there is unfortunately no satisfactory classification result for subalgebras\footnote{Conjugacy classes of subalgebras of the \sch algebra are classified in \cite{Winternitz}.}. However, in addition to being a subalgebra, there are various physically motivated conditions, that are naturally imposed upon $\mathfrak{h}$:

\medskip
\noindent
{\bf Assumption 1 (No translation condition).}  $\mathfrak{h}$ does not contain $P^i$.
\medskip

This is natural because $P^i$ will be realized as infinitesimal translations in the geometry, and should not be included in the stabilizer of a point on the homogeneous space $G/H$.
Another condition we impose is:

\medskip
\noindent
{\bf Assumption 2 (Lorentz subgroup condition).} $\mathfrak{h}$ contains $J^{ij}$ and $G^i$.
\medskip

This condition is needed because we want to respect $d$-dimensional local Lorentz symmetry, 
which is crucial for the equivalence principle of general relativity\footnote{In the literature, stronger constraints are imposed on $G$ \cite{Weinberg1}, although for our purposes the Lorentz subgroup condition is strong enough. }.

\medskip

Although our methods apply to \sch cosets in arbitrary dimensions, let us concentrate on the case of $\textrm{dim}\,G/H=5$ and $\textrm{dim}\,H=4$. This is the case discussed recently in the literature, which in the context of the non-relativistic AdS/CFT correspondence is conjectured to be dual to $(2+1)$-dimensional non-relativistic conformal field theories \cite{Son:2008ye, Balasubramanian:2008dm}.

If we impose the above two assumptions then $\mathfrak{h}$ is spanned by $J,G^1,G^2$ and one more generator, and the possible choices are
\be
\ba
\mathfrak{h}_{(1)} 	&= \langle  J, G^1, G^2, \alpha C + \beta M + \gamma D  \rangle 
\qquad (\alpha\not=0)\,,	\cr
\mathfrak{h}_{(2)} 	&= \langle  J, G^1, G^2, \beta M+ \gamma D  \rangle	
\quad (\beta\not=0)\, , \\
\mathfrak{h}_{(3)}      &= \langle J, G^2, G^2, D \rangle.
\ea
\ee

The Ad$(H)$-invariant two-forms are obtained by solving for $\Omega$ in (\ref{TwoForm}). This requires in particular a specification of the basis of generators of the complement  $\mathfrak{m}_i$ of each subalgebra. 
Define
\be
\ba
\mathfrak{m}_{(1)} &= \langle H, P^1, P^2, M, D  \rangle\, , \cr
\mathfrak{m}_{(2)} &= \langle H, P^1, P^2, C, D  \rangle\, , \cr
\mathfrak{m}_{(3)} &= \langle H, P^1, P^2, C, M  \rangle\, , \cr
\ea
\ee

Let us consider in detail the case $\mathfrak{h}_{(1)}$. 
Assuming that $\beta\not=0$,
the structure constants relevant for (\ref{TwoForm}) are
\be
\ba
f_{[i] J}{}^{[j]} &= \left(
			\begin{array}{ccccc}
			0&0&0&0&0\cr
			0&0&+1&0&0\cr
			0&-1&0&0&0\cr
			0&0&0&0&0\cr
			0&0&0&0&0
			\end{array}
			\right)
\,,\quad
f_{[i] \alpha C + \beta M + \gamma D}{}^{[j]} = \left(
			\begin{array}{ccccc}
			-2\gamma&0&0&0&-\alpha\cr
			0&-\gamma&0&0&0\cr
			0&0&-\gamma&0&0\cr
			0&0&0&0&0\cr
			0&0&0&2 \beta&2\gamma
			\end{array}
			\right)	,
\cr
f_{[i] G^1}{}^{[j]} &= \left(
			\begin{array}{ccccc}
			0&+1&0&0&0\cr
			0&0&0&1&0\cr
			0&0&0&0&0\cr
			0&0&0&0&0\cr
			0&0&0&0&0
			\end{array}
			\right)	
\,,\qquad\quad \ 
f_{[i] G^2}{}^{[j]} = \left(
			\begin{array}{ccccc}
			0&0&+1&0&0\cr
			0&0&0&0&0\cr
			0&0&0&1&0\cr
			0&0&0&0&0\cr
			0&0&0&0&0
			\end{array}
			\right)	\,.								
\ea
\ee
Then solving for $\Omega$ in (\ref{TwoForm}) yields
for $\gamma\not=0$ the resulting two-form is degenerate. 
For $\alpha, \beta\not=0$ and $\gamma=0$ there exists a non-degenerate two-form 
\be
\Omega^{(1)}_{[i][j]} = \left(
			\begin{array}{ccccc}
			 \Omega_{HH} & 0 & 0 & -\Omega_{PP}& 0 \cr
			0&\Omega_{PP}&0&0&0\cr
			0&0&\Omega_{PP}&0&0\cr
			-\Omega_{PP}&0&0&0&0\cr
			0&0&0&0&-2 {\beta\over\alpha}\Omega_{PP}
			\end{array}
			\right)\,,  \qquad \alpha, \beta\not=0\,,\ \gamma=0\,.
\label{eq.Omega1}
\ee
Parameterizing the coset elements as
\be
g=e^{x_H H} e^{x_M M} e^{x_i P^i} e^{x_D D}\,,
\ee
then the vielbeine are given by
\be
e_H=e^{2 x_D} dx_H, \quad e_M=dx_M, \quad e_i=e^{x_D} dx^i\,, \quad 
e_D=dx_D \,.
\ee
One may take $\Omega_{PP}=1$ by using the overall scaling. Then $\sigma:=\Omega_{HH}$ can be freely chosen, 
and by suitable coordinate transformation, we can set $2 \beta/\alpha =1$, which yields 
\be
ds^2=r^2(-2 dx^+ dx^- +dx^i dx^i)+\frac{dr^2}{r^2}- \sigma r^4 (dx^+)^2,
\label{eq.Son}
\ee
where we defined 
\be
e^{x_D}=r, \quad x_H=x^+, \quad x_M=x^- \,.
\ee
For $\sigma=0$, this is the DLCQ of AdS$_5$ \cite{Goldberger:2008vg,Barbon:2008bg}. Although the metric looks locally the same as that of AdS$_5$, the presence of $M$, which commutes with all other generators, means that the eigenvalue of $M$ is quantized and the corresponding direction, namely the $x^+$-direction, is compactified.
The deformation term proportional to $\sigma$ is nothing but the deformation term that was also observed to be present in \cite{Son:2008ye, Balasubramanian:2008dm}. One way of understanding the appearance of this deformation term is the null Melvin twist \cite{Herzog:2008wg,Maldacena:2008wh,Adams:2008wt}, but from the viewpoint of the coset this is simply a deformation parameter of the invariant two-form\footnote{Applying this method to the pp-wave case \cite{Metsaev}, which is reductive, would  also yield a deformation term. However, this can be removed by a coordinate transformation and has no physical significance.}.

This coset, constructed from $\mathfrak{h}_{(1)}=\{J,G^i,\alpha C+\beta M\} \, (\alpha, \beta\ne 0)$ is an interesting example of a non-reductive coset. 
 The examples of non-reductive cosets are scarce in the literature, and in four dimensions and less. A classification of Lorentzian non-reductive homogeneous spaces in four and less dimensions appears in \cite{FelsRenner}.

In a similar fashion, one can analyze the case of $\mathfrak{h}_{(2)}$ and $\mathfrak{h}_{(3)}$, and in both instances we have verified the non-existence of a non-degenerate invariant two-forms. This implies that under the above assumptions, the metric \eqref{eq.Son} is unique:

\medskip
{\bf Uniqueness.} Under the Assumptions 1 and 2 above, the 5d coset of $\widetilde{\mathfrak{Sch}_2}$ is unique, and the metric is given by \eqref{eq.Son}.
\medskip

One way to escape this uniqueness theorem is to abandon assumption 2. 
Although in such cases the rotation symmetry or Galilean boost symmetry is broken as a symmetry of the local frames, they still exist as symmetries of the background, and these can potentially become useful in the future study of non-relativistic AdS/CFT correspondence.

Relaxing assumption 2, there are two further choices for subalgebras:
\be
\ba
\mathfrak{h}_{(4)} 	&= \langle  G^1, G^2, C, D+\alpha M  \rangle\,,	\cr
\mathfrak{h}_{(5)} 	&= \langle   H, C, D   \rangle \oplus \langle  \alpha J + \beta M  \rangle	\,.
\ea
\ee
Their complements can be chosen as
\be
\ba
\mathfrak{m}_{(4)} &= \langle H, J,  P^1, P^2, M \rangle\,, \cr
\mathfrak{m}_{(5)} &= \langle P^1, P^2, G^1, G^2, J  \rangle \,,
\ea
\ee
and then straightforward computation shows that for $\mathfrak{h}_{(4)}$ there do not exist any non-degenerate two-forms. 
For $\mathfrak{h}_{(5)}$ with $\beta\ne 0$, we obtain
\be
\Omega =2 \Omega_{P_1,G_2} (e_{P_1} e_{G_2}-e_{P_2} e_{G_1})+\Omega_{JJ} e_J^2\,.
\label{eq.Omega4}
\ee
Parameterizing the coset elements as
\be
g=e^{x_i P_i} e^{y_i G_i} e^{x_J J},
\ee
the invariant one-forms are
\be
\ba
e_{P_1}&=dx_1 \cosh x_J-dx_2 \sinh x_J ,\quad 
e_{P_2} &=dx_1 \sinh x_J+dx_2 \cosh x_J ,\cr
e_{G_1}&=dy_1 \cosh x_J-dy_2 \sinh x_J ,\quad 
e_{G_2}&=dy_1 \sinh x_J+dy_2 \cosh x_J ,\cr
e_J&=-\gamma (dx_1 y_1+dx_2 y_2)+dx_J\, ,
\label{eq.vielcase4}
\ea
\ee
where $\gamma=\frac{\alpha}{\beta}$.
Equations \eqref{eq.Omega4} and \eqref{eq.vielcase4} yields a metric with \sch symmetry:
\be
ds^2=(dx_1 dy_2-dx_2 dy_1) \cosh 2 x_J + \sigma (dx_J-\gamma(y_1 dx_1 +y_2 dx_2))^2
\ee
with $\sigma\ne 0$. Unfortunately, the signature of this spacetime is $(2,3)$, and as such does not seem to be suitable for applications in AdS/CFT. Therefore, even by relaxing the conditions in assumption 2, the background (\ref{eq.Son}) seems to be unique.


\subsection{The Cases of $z\ne 2$}

In a  non-relativistic spacetime, we can scale time and space differently:
\be
t\to \lambda^z t, \quad x\to \lambda x \,,
\ee
where the parameter $z$ is called the dynamical exponent. The discussion so far corresponds to the case $z=2$, and we are now going to consider the case with arbitrary dynamical exponent $z \neq 2$. The algebra (which we call $\widetilde{\mathfrak{Sch}}_{d,z}$) is given by
\begin{eqnarray}
&& [J^{ij},J^{kl}]=-\delta^{ik} J^{jl}+\delta^{il} J^{jk} 
- \delta^{jl}J^{ik} + \delta^{jk} J^{il}\,, \nn  \\  
&& [J^{ij},P^k] = -\delta^{ik}P^j + \delta^{jk}P^i\,, \qquad 
[J^{ij},G^k] = -\delta^{ik} G^j + \delta^{jk}G^i\,, \nn \\ 
&& [H,G^i]=P^i\,, \qquad [P^i,G^j] = \delta^{ij}M\,, \qquad [D^z,H]=zH\,, \nn \\ 
&& [D^z,P^i] =P^i\,, \qquad [D^z,G^i] = (1-z) G^i\,, \qquad 
[D^z,M] = (2-z) M\,. \label{nini}
\end{eqnarray}
Note that $C$ is broken in the case with $z\neq 2$\,.

Under the two assumptions of the previous section, a natural coset candidate is
\begin{eqnarray}
\mathfrak{h}=\{G^i,J^{ij}\}\,, \quad \mathfrak{m} = \{H,M,P^i,D^z\}\,.
\end{eqnarray} 
The invariant two-form associated to this choice is
\be
\Omega=
\left(
\begin{array}{lllll}
\Omega_{HH} & -\Omega_{PP} &0 & 0 &\Omega_{HD} \\ 
-\Omega_{PP} & 0 &0&0 &0\\
0 &0 &\Omega_{PP} &0 &0 \\
0& 0& 0& \Omega_{PP}& 0\\
\Omega_{DH}&0&0&0& \Omega_{DD}
\end{array}
\right).
\ee 
A group
element of $G/H$ is represented by $g=\e^{x_H H} \e^{x_M M} \e^{x_i P^i}
\e^{x_D D^z}$, and the vielbeine are
\begin{eqnarray}
e_H = \e^{z x_D} dx_H\,, \qquad e_M = \e^{(2-z)x_D}dx_M\,, 
\qquad e_i = \e^{x_D}dx_i\,, \qquad e_D = dx_D\,.
\end{eqnarray}
The metric is, up to coordinate transformations, given by
\begin{eqnarray}
ds^2 &=& -2 e_He_M + e_i^2 + e_D^2 + \sigma e_H^2\nn \\ 
&=& r^2(-2dx^+dx^- + dx^idx^i) + \frac{dr^2}{r^2}+\sigma r^{2z}(dx^+)^2\,, 
\end{eqnarray}
where we have identified as $x_H = x^+, x_M = x^-, \e^{x_D}=r$\,.
When $\sigma=0$, this again yields the DLCQ of AdS, and $\sigma$ is a deformation term \cite{Balasubramanian:2008dm} similar to the one discussed in the previous section.


\subsection{A comment on super-cosets} 

It would be interesting to consider the super-cosets related to non-relativistic AdS/CFT backgrounds. 
The most interesting example of super coset is represented by a subalgebra of psu(2,2$|$4) \cite{Sakaguchi:2008rx,Sakaguchi:2008ku}. This symmetry is known to be realized by the background consisting of the metric of  DLCQ of AdS$_5$ times S$^5$. That is, the  $x^-$-compactification breaks the relativistic conformal symmetry to the Schr\"odinger and the 16 superconformal symmetries are broken to 8. 

For 24 supercharges there should be no deformation term \cite{Maldacena:2008wh}\footnote{In \cite{HY} a class of supersymmetric Schr\"odinger backgrounds is discussed, however, these are not homogeneous spaces.}. 
However, it seems difficult to see this from the coset description, since the argument for the metric is unaltered.
This result is not so surprising since the on-shell condition of supergravity is not 
taken into account, as well as the fact that  the number of supersymmetries is not 
maximal. In particular, the presence of the B-field, which breaks the symmetry of the metric,  is not included in our argument.


\section{Gravity Dual of Lifshitz Fixed Points and Cosets}
\label{sec:Lif}

In the previous section we fix $\mathfrak{g}$ to be $\widetilde{\mathfrak{Sch}}$ and considered various choices of subgroups $\mathfrak{h}$. Consider now cosets, where $\mathfrak{g}$ is a subalgebra of the Schr\"odinger algebra.

First we need to address the question of which subgroup of $\widetilde{\mathfrak{Sch}}$ we should take as $\mathfrak{g}$.  Let us first discuss the case $z=2$. One possibility is to search for an interesting subgroup which does not contain $M$. This is because in the discussion above $M$ corresponds to a $x^{+}$-directions, and this is the origin of the difficulties associated with DLCQ in the dual CFT.

Since $[P^i,G^j]=\delta^{ij} M$, we have to remove either $P^i$ or $G^i$. Since we want to keep translation invariance, let us remove $G^i$. Then again since $[C,P^i]=G^i$, we also need to remove $C$. The the remaining generators $H,D,P^i,J$ span a subgroup of  $\widetilde{\mathfrak{Sch}}$. In the case $z\ne 2$ \eqref{nini}, we can consider the same algebra.
We thus consider $\mathfrak{g}=\langle J^{ij},P^i,H,D^z,M\rangle$.
Consider the case $d=2$, since higher dimensional case are similar. If we are going to consider a 4d coset of $\mathfrak{g}$, $\mathfrak{h}$ is one-dimensional and assumption 2 above uniquely determines $\mathfrak{h}$ to be $\mathfrak{h}= \langle J^{ij}\rangle$ \footnote{In this case, Galilean symmetry is not inlucded in $\mathfrak{h}$ since it is broken from the outset. However, the theory discussed here has a 
``doubled'' Galilean symmetry acting on particles and anti-particles in an opposite way.}.
Then the relevant commutation relations are
\be
[J, P^1] = - P^2 \,,\qquad 
[J, P^2]= P^1 \,,\qquad [D^z, H] = z H \,,\qquad 
[D^z, P^i] = P^i \,.
\ee
Let $\mathfrak{m} = \langle H, P^1, P^2, D^z\rangle$. Then solving for (\ref{TwoForm}) we obtain
\be
\Omega^{}_{[i][j]}=
\left(
\begin{array}{llll}
 \Omega_{HH} & 0 & 0 &  \Omega_{HD} \\
  0& \Omega_{P^1 P^1} & 0 &  0\\
   0& 0 & \Omega_{P^1 P^1} &  0 \\
    \Omega_{HD} & 0 & 0 &  \Omega_{DD} \\
\end{array}
\right).
\ee
In this case, all the deformation parameters in the invariant two-forms are removed by coordinate transformations, and the resulting metric is 
\be
ds^2 = - r^{2z} dt^2 + r^2 dx_i^2 + {dr^2 \over r^2}.
\ee
This is precisely the background in \cite{Kachru:2008yh}, which is the candidate gravity dual of the Lifshitz fixed point. Again, similar arguments as in the previous section seem to show that this
is the unique 4d coset of this group even when the assumption 2 is relaxed.

\newpage


\section*{Acknowledgements}

We would like to thank  T. Azeyanagi, N. Bobev, S. Detournay, 
T. Dimofte, M. Fels, J. Gomis, A. Mikhailov, M. Mulligan, H. Ooguri and M. Sakaguchi
for useful comments and discussions. We thank the KITP, Santa Barbara, for hospitality during some of this work. 
This work is supported by DOE grant DE-FG03-92-ER40701 (SSN and MY), 
 by a Caltech John A. McCone Postdoctoral Fellowship (SSN),  
by the World Premier International Research Center Initiative, MEXT, Japan,
by the JSPS fellowships for Young Scientists, and by Global COE Program ``the Physical Sciences Frontier", MEXT, Japan (MY), 
and by the Grant-in-Aid for the Global COE Program ``The Next Generation of Physics, Spun from Universality and Emergence", MEXT, Japan (KY).


\bibliographystyle{JHEP}
\renewcommand{\refname}{Bibliography}
\parskip=-3pt


\providecommand{\href}[2]{#2}\begingroup\raggedright\endgroup


\end{document}